\documentclass[prd,superscriptaddress,nofootinbib,showpacs,preprint]{revtex4}
\usepackage{graphicx}
\usepackage[usenames,dvipsnames]{color}
\usepackage{amsmath,amssymb}
\usepackage[colorlinks]{hyperref}

\newcommand{\be}{\begin{equation}}
\newcommand{\ee}{\end{equation}}
\begin{document}
\title{TeV scale Left Right Symmetry with spontaneous \\ D-parity breaking}

\author{Debasish Borah}
\email{debasish@phy.iitb.ac.in}
\affiliation{Indian Institute of Technology Bombay, Mumbai-400076, India}
\author{Sudhanwa Patra}
\email{sudhakar@prl.res.in}
\affiliation{Physical Research Laboratory, Ahmedabad-380009, India}
\author{Utpal Sarkar}
\email{utpal@prl.res.in}
\affiliation{Physical Research Laboratory, Ahmedabad-380009, India}

\begin{abstract}
The different scenarios of spontaneous breaking of D-parity have been studied in both
non-supersymmetric and supersymmetric version of the left-right symmetric models(LRSM). We explore the
possibility of a TeV scale $SU(2)_R$ breaking scale $M_R$ and hence TeV scale right
handed neutrinos from both minimization of the scalar potential as well as the coupling
constant unification point of view. We show that although minimization of the scalar
potential allows the possibility of a TeV scale $M_R$ and tiny neutrino masses in LRSM
with spontaneous D-parity breaking, the gauge coupling unification at a high scale
$\sim 10^{16}$ GeV does not favour a TeV scale symmetry breaking except
in the supersymmetric left-right (SUSYLR) model with Higgs doublet and bidoublet. The phenomenology of neutrino mass is also discussed.
\end{abstract}

\pacs{12.60.Fr,12.60.Jv,12.60.-i,14.60.Pq,14.60.St}
\maketitle

\section{Introduction}
\label{sec:intro}
Left-Right symmetric model(LRSM) is a novel extension of the standard model
of particle physics \cite{Pati:1974yy,Mohapatra:1974gc, Senjanovic:1975rk,
Mohapatra:1980qe, Deshpande:1990ip}. In such models the parity is spontaneously
broken and the smallness of neutrino masses \cite{Fukuda:2001nk, Ahmad:2002jz,
Ahmad:2002ka, Bahcall:2004mz} arises in a natural way via seesaw mechanism
\cite{Minkowski:1977sc, GellMann:1980vs, Yanagida:1979as, Mohapatra:1979ia}. Incorporating supersymmetry(susy) into such models comes with couple of other advantages in terms of the gauge hierarchy problem, coupling constant unification among many others. Another advantage in such susy models is that they provide a natural candidate for dark matter in terms of the lightest super-particle (LSP). In the minimal supersymmetri standard model(MSSM), this LSP is stable only if we incorporate an extra symmetry called R-parity $R_p = (-1)^{3(B-L)+2s}$. However in supersymmetric left right (SUSYLR) models \cite{Aulakh:1997ba, Aulakh:1997fq, Aulakh:1998nn, Aulakh:1997vc} based on the gauge group $SU(3)_C\times SU(2)_L \times SU(2)_R \times U(1)_{B-L}$ this R-parity is a part of the gauge symmetry and hence need not be put by hand. Since $U(1)_{B-L}$ symmetry is broken by a Higgs triplet with even $B-L$ quantum number, R-parity is still preserved at low energy.

In the usual LRSM, the scale of parity breaking and $SU(2)_R$ gauge symmetry
breaking are identical which is not necessary. There have been lots of studies on left-right symmetric models where the parity symmetry gets broken much before the $SU(2)_R$ gauge symmetry breaks by so called spontaneous D-parity breaking \cite{Chang:1983fu,Chang:1984uy}. In this paper we analyses various types of susy and non-susy left-right models with spontaneous D-parity breaking and check analytically whether the minimization of the scalar potential allows a
TeV scale $SU(2)_R$ breaking scale (provided parity breaks at much higher scale) as well as tiny neutrino masses. We then check whether such a choice
of intermediate symmetry breaking scales unifies the gauge coupling constants in
the SUSYLR framework. We discuss the possible phenomenology
of neutrino mass in each cases separately.

\textit{Motivation and Outlook}: Since many papers exist in the literature studying these aspects of the left-right symmetric models, we summarize here our motivation for this study and how our analysis differs from earlier works. Before the precision measurements of the weak mixing angle and the strong coupling constants, the evolution of the gauge coupling constants could allow low-scale left-right symmetry breaking \cite{Rizzo:1981su}. This could be achieved with a single stage symmetry breaking. Later it was found that by invoking more intermediate scales, it is possible to have more freedom to adjust the different symmetry breaking scales. However, after the precision electroweak measurements at LEP, it was found that the simplest left-right symmetric models would not allow a left-right symmetry breaking below $10^{12}$ GeV, in both single stage symmetry breaking as well as multi-stage symmetry breaking \cite{Langacker:1991an,Langacker:1991pg,Langacker:1991zr}. $SO(10)$ based models also got constrained with the allowed intermediate scale in the range of $10^9-10^{10}$ GeV \cite{Deshpande:1992au,Deshpande:1992em}. Introducing the Pati-Salam symmetry breaking scale would not allow lowering the left-right symmetry breaking scale both in the supersymmetric as well as the non-supersymmetric models. It would be possible to break the $SU(2)_R$ to $U(1)_R$ at a higher scale and then break the group $U(1)_R$ at a lower scale, but the breaking scale of $SU(2)_R$ could not be lowered, keeping the theory consistent with the potential minimization and gauge coupling evolution. 

In a recent paper, it has been demonstrated that by introducing additional scalars it is possible to lower the scale of left-right symmetry breaking, i.e., break the symmetry group $SU(2)_R$ \cite{Dev:2009aw}. In this paper we studied the different symmetry breaking patterns to check the consistency with the potential minimization and gauge coupling evolution and see which of these models could allow TeV scale left-right symmetry breaking. We restricted our analysis to only a single stage symmetry breaking, because by introducing the additional symmetry breaking scales it was not found to help lowering the left-right symmetry breaking scales. Of course, our analysis does not rule out other possibilities of lowering the left-right breaking scale by introducing newer symmetry breaking scales and new physics. However, this analysis demonstrates, within the simplest framework of single stage symmetry breaking, which models are consistent with potential minimization, gauge coupling unification, and allows a TeV scale left-right symmetry breaking. 

\indent This paper is organized as follows. In the next section [\ref{sec2}], we will study the potential minimization of non-susy and susy version of various left-right symmetric models and check the minimization of the scalar potential. Then in section [\ref{sec:gauge}]
we study the gauge coupling unification in all the SUSYLR models we have considered
and discuss the neutrino mass in sections [\ref{subsec:neumass}] and [\ref{numass:triplet}]. We discuss the results and conclusion in section [\ref{results}] and finally conclude in section [\ref{conclusion}].

\section{LR models with spontaneous D-parity breaking}
\label{sec2}
In left
right symmetric models with spontaneous D-parity breaking, the discrete parity
symmetry gets broken (by the vev of a parity odd singlet scalar field) much before the
$SU(2)_R$ gauge symmetry breaks. The gauge group is
effectively $SU(3)_C\times SU(2)_L \times SU(2)_R \times U(1)_{B-L} \times P$, where
$P$ is the discrete left-right symmetry which we call D-parity. This D-parity symmetry is different from the Lorentz parity in the sense that Lorentz parity interchanges left handed fermions with the right handed ones but the bosonic fields remain the same. Whereas, the D-parity also interchanges the $SU(2)_L$ Higgs fields with the $SU(2)_R$ Higgs fields. The parity odd singlet field breaks this gauge symmetry at high
scale $\sim (10^{16}-10^{19})$ GeV to $SU(3)_C\times SU(2)_L \times SU(2)_R \times U(1)_{B-L}$ which further
breaks down to the standard model gauge group $SU(3)_C \times SU(2)_L \times U(1)_Y$ at a lower scale. The D-parity breaking introduces an asymmetry between left and right handed Higgs fields and makes the coupling constants of $SU(2)_R$ and $SU(2)_L$ evolve separately under the renormalization group. It should be noted that this D-parity breaking is different from the low energy parity breaking observed in the weak interactions which arises as a result of $SU(2)_R$ gauge symmetry breaking at a scale higher than the electroweak scale. In such D-parity breaking scenario the seesaw relation also gets modified from usual LRSM. Although the type I seesaw term
still remains sensitive to the  $SU(2)_R$ breaking scale $M_R$, the other seesaw terms
namely type II and type III \cite{Foot:1988aq} becomes sensitive to the D-parity breaking
scale. A very high value of parity breaking scale therefore leads to type I seesaw dominance. In this section we are going to discuss various such models with different particle contents.

\subsection{LRSM with Higgs doublets}
We first study the non-Susy left-right symmetric extension of the standard
model with only Higgs doublets. In addition to the usual fermions of the
standard model, we require the right-handed neutrinos to complete the
representations. One of the important features of the model is that it
allows spontaneous parity violation. The Higgs representations then
requires a bi-doublet field, which breaks the electroweak symmetry and
gives masses to the fermions. But the neutrinos can have a Dirac mass
only, which is then expected to be of the order of other fermion masses.
To implement the see-saw mechanism and obtain the observed tiny mass
of the left-handed neutrinos naturally, one also introduces a singlet fermion 
plus fermion triplet. However, we shall restrict ourselves to the scalar sector and shall not
discuss the implications of the singlet neutrinos and the neutrino
masses.

The particle content of the Left-Right symmetric model with Higgs doublet is
$${\rm Fermions:}~~  Q_L \equiv (3,2,1,1/3),~~  Q_R \equiv (3,1,2,1/3), ~~
\Psi_L \equiv (1,2,1,-1),~~  \Psi_R \equiv (1,1,2,-1) $$
$${\rm Scalars:} \quad \Phi \equiv (1,2,2,0), \quad H_L \equiv (1,2,1,1),
\quad H_R \equiv (1,1,2,1) \quad \rho \equiv (1,1,1,0) $$
where the numbers in the brackets are the quantum numbers corresponding to the
gauge group $SU(3)_C \times SU(2)_L \times SU(2)_R \times U(1)_{B-L} $.
In addition to the bi-doublet scalar field $\Phi$, we also introduced two
doublet fields $H_L$ and $H_R$ to break the left-right symmetry and contribute
to the neutrino masses. The scalar singlet $\rho$ is a D-parity odd field and
changes sign under the exchange of $SU(2)_L$ with $SU(2)_R$. Thus the
symmetry breaking pattern becomes
$$ SU(2)_L \times SU(2)_R \times U(1)_{B-L} \times P \quad \underrightarrow{\langle \rho \rangle}\quad SU(2)_L \times SU(2)_R \times U(1)_{B-L} $$
$$ \underrightarrow{\langle
H_R \rangle} \quad SU(2)_L\times U(1)_Y \quad \underrightarrow{\langle \Phi \rangle} \quad U(1)_{em}$$
We denoted the vacuum expectation values of the neutral components of the Higgs fields as
$$ \langle \Phi_1 \rangle = v_1,v_2, \quad \langle H_L \rangle = v_L,\quad \langle H_R \rangle =v_R, \quad  \langle \rho \rangle = s$$
The scalar potential with all these fields can then be written as
\begin{eqnarray*}
\lefteqn{V= \mu_1^2 \text{Tr}[\Phi_1^{\dagger}\Phi_1]+\mu_2^2 (\text{Tr}[\Phi_2 \Phi_1^{\dagger}]+
\text{Tr}[\Phi_2^{\dagger}\Phi_1] )+ \lambda_1 (\text{Tr}[\Phi_1^{\dagger} \Phi_1])^2+
\lambda_2[(\text{Tr}[\Phi_2 \Phi_1^{\dagger}])^2+(\text{Tr}[\Phi_2^{\dagger}\Phi_1 ])^2 ]}
\nonumber \\
&& +\lambda_3 \text{Tr}[\Phi_2 \Phi_1^{\dagger}]\text{Tr}[\Phi_2^{\dagger}\Phi_1 ]+
\lambda_4 \text{Tr}[\Phi_1^{\dagger}\Phi_1](\text{Tr}[\Phi_2 \Phi_1^{\dagger}]+
\text{Tr}[\Phi_2^{\dagger}\Phi_1] )+\mu^2_h (H^{\dagger}_L H_L+H^{\dagger}_R H_R)
\nonumber \\
&& + \lambda_5 [(H^{\dagger}_L H_L)^2+(H^{\dagger}_R H_R)^2]+\lambda_6 (H^{\dagger}_L H_L)
(H^{\dagger}_R H_R)+\alpha_1 \text{Tr}[\Phi_1^{\dagger}\Phi_1](H^{\dagger}_L H_L+ H^{\dagger}_R H_R)
\nonumber \\
&& +\alpha_2 (H^{\dagger}_L\Phi_1\Phi_1^{\dagger} H_L+H^{\dagger}_R\Phi_1^{\dagger}\Phi_1 H_R)+
\alpha_3 (H^{\dagger}_L\Phi_2\Phi_2^{\dagger} H_L+H^{\dagger}_R\Phi_2^{\dagger}\Phi_2 H_R)+
\alpha_4 (H^{\dagger}_L\Phi_1\Phi_2^{\dagger} H_L
\nonumber \\
&& +H^{\dagger}_R\Phi_1^{\dagger}\Phi_2 H_R)+\alpha^*_4 (H^{\dagger}_L\Phi_2\Phi_1^{\dagger} H_L+
H^{\dagger}_R\Phi_2^{\dagger}\Phi_2 H_R) + \mu_{h\phi1} (H^{\dagger}_L \Phi_1 H_R + H^{\dagger}_R
\Phi^{\dagger}_1 H_L)+\mu_{h\phi2} (H^{\dagger}_L \Phi_2 H_R
\nonumber \\
&& + H^{\dagger}_R \Phi^{\dagger}_2 H_L)-\mu^2_{\rho} \rho^2 +\lambda_7 \rho^4 +
M \rho(H^{\dagger}_L H_L -H^{\dagger}_R H_R)+\lambda_8 \rho^2(H^{\dagger}_L H_L+
H^{\dagger}_R H_R)
\nonumber \\
&& \lambda_9 \rho^2 \text{Tr}[\Phi_1^{\dagger}\Phi_1]+\lambda_{10}\rho^2
[\text{Det}[\Phi_1 ]+\text{Det}[\Phi^{\dagger}_1]]
\end{eqnarray*}
where $ \Phi_2 = \tau_2 \Phi^*_1 \tau_2 $.

To find a consistent solution we now minimize the scalar potential and obtain
\begin{eqnarray}
\frac{\partial V}{\partial v_L} = \mu^2_L v_L + \lambda_5 v^3_L+
\frac{\lambda_6}{2}v_L v^2_R+ \mu_{h\phi} (v_1+v_2)v_R = 0
\label{eq1}  \\
\frac{\partial V}{\partial v_R} = \mu^2_R v_R + \lambda_5 v^3_R+
\frac{\lambda_6}{2}v_R v^2_L +\mu_{h\phi} (v_1+v_2)v_L = 0
\label{eq2}
\end{eqnarray}
where $\mu^2_L$ and $\mu^2_R$ are effective mass terms of $ H_L$ and $H_R$ given by
\begin{eqnarray}
\mu^2_L = \mu^2_h +M s + \lambda_8 s^2 + (\alpha_4+\alpha^*_4) v_1v_2+ \alpha_1
(v^2_1+v^2_2)+\alpha_2 v^2_2 +\alpha_3 v^2_1  \nonumber \\
\mu^2_R = \mu^2_h -M s + \lambda_8 s^2 + (\alpha_4+\alpha^*_4) v_1v_2+ \alpha_1
(v^2_1+v^2_2)+\alpha_2 v^2_2 +\alpha_3 v^2_1
\label{mr}
\end{eqnarray}
Thus after the singlet field $\eta$ gets a vev the left handed Higgs doublet becomes
heavy and decouple whereas the right handed Higgs can be much lighter by appropriate
fine tuning of the parameters in (\ref{mr}). From equations (\ref{eq1}), (\ref{eq2}) we get
\begin{equation*}
v_Lv_R (2 M s) + (\lambda_5 -\frac{\lambda_6}{2})(v^2_L-v^2_R)v_Lv_R+\mu_{h\phi} (v_1+v_2)(v^2_R-v^2_L) = 0
\end{equation*}
Thus a non-zero value of $ \langle \rho \rangle = s$ does not allow a solution with
$v_L = v_R$. The seesaw relation from the above equation is
\begin{equation*}
v_Lv_R = \frac{\mu_{h\phi}(v_1+v_2)(v^2_L-v^2_R)}{2M s + (\lambda_5-\frac{\lambda_6}{2})(v^2_L-v^2_R)}
\end{equation*}
Assuming $v_L \ll v_R \ll s, M $ will give
\begin{equation}
v_L = \frac{-\mu_{h\phi} (v_1+v_2) v_R}{2 M s}
\end{equation}
Thus we can have small $v_L/v_R $ by appropriately choosing the scales of
$M, s, \mu_{h\phi}$ which will account for tiny neutrino masses. In contrast
LRSM without D-parity breaking where the right handed scale $v_R$ has to be
very high to account for small $v_L/v_R $, here we can have $v_R$ of TeV
scale also. For example, if we set $\mu_{h\phi}= M = s = 10^{8} $ GeV, and $v_{1,2} \sim M_Z $
then $\frac{v_L}{v_R}$ comes out to be of the order $10^{-6}$ which is desired for
type III seesaw to dominate as we will see when we discuss neutrino masses.
The gauge coupling unification has been studied extensively in this model, so
we shall not repeat them here. In the absence of D-parity breaking the left-right
symmetry breaking scale comes out to be very high, but in D-parity violating
models it is possible to lower the scale of left-right symmetry breaking with
some amount of fine tuning of parameters. However, for the supersymmetric
models restrictions are more stringent, so we shall study them in details.

\subsection{LRSM with Higgs triplets}
In this section we shall study the left-right symmetric models with a
different particle contents. The usual fermions, including the right-handed
neutrinos, belong to the similar
representations as in the previous section. However the scalar sector now
contains triplet Higgs scalars in addition to the bi-doublet Higgs scalar to
break the left-right symmetry. The triplet Higgs scalars can then give
Majorana masses to the neutrinos and allow seesaw mechanism without the
need for any additional singlet fermions. The parity odd singlet scalar was
originally introduced in this model, so we shall include them in our
discussions.

The particle content of LRSM with Higgs triplets is
$${\rm Fermions:}~~  Q_L \equiv (3,2,1,1/3),~~  Q_R \equiv (3,1,2,1/3), ~~
\Psi_L \equiv (1,2,1,-1),~~  \Psi_R \equiv (1,1,2,-1) $$
$${\rm Scalars:} \quad
\Phi \equiv (1,2,2,0), \quad \triangle_L \equiv (1,3,1,2), \quad \triangle_R \equiv (1,1,3,2)
\quad \rho \equiv (1,1,1,0) $$
The symmetry breaking pattern in this model remains the same as in the previous
model although the structure of neutrino masses changes. In the symmetry breaking
pattern, the scalar $\Delta_c$ now replaces the role of $H_R$, but otherwise
there is no change. The vacuum expectation
values of the neutral components of the Higgs fields are denoted by
$\Phi_1, \triangle_L, \triangle_R, \rho$ as
$$ \langle \Phi_1 \rangle = v_1,v_2, \quad \langle \triangle_L \rangle = v_L, 
\quad \langle \triangle_R \rangle =v_R, \quad  \langle \rho \rangle = s$$
The scalar potential can then be written as
\begin{eqnarray*}
\lefteqn{V= \mu_1^2 \text{Tr}[\Phi_1^{\dagger}\Phi_1]+\mu_2^2 (\text{Tr}[\Phi_2 \Phi_1^{\dagger}]+
\text{Tr}[\Phi_2^{\dagger}\Phi_1] )+ \lambda_1 (\text{Tr}[\Phi_1^{\dagger} \Phi_1])^2+
\lambda_2[(\text{Tr}[\Phi_2 \Phi_1^{\dagger}])^2+(\text{Tr}[\Phi_2^{\dagger}\Phi_1 ])^2 ]}
\nonumber \\
&& +\lambda_3 \text{Tr}[\Phi_2 \Phi_1^{\dagger}]\text{Tr}[\Phi_2^{\dagger}\Phi_1 ]+
\lambda_4 \text{Tr}[\Phi_1^{\dagger}\Phi_1](\text{Tr}[\Phi_2 \Phi_1^{\dagger}]+
\text{Tr}[\Phi_2^{\dagger}\Phi_1] )+\mu^2_{\triangle} (\text{Tr}[\triangle^{\dagger}_L
\triangle_L ]+\text{Tr}[\triangle^{\dagger}_R \triangle_R ])
\nonumber \\
&& +f_1 [(\text{Tr}[\triangle^{\dagger}_L \triangle_L ])^2 +(\text{Tr}[\triangle^{\dagger}_R
\triangle_R ])^2 ]+f_2 (\text{Tr}[\triangle_L \triangle_L ]\text{Tr}[\triangle^{\dagger}_L
\triangle^{\dagger}_L ]+\text{Tr}[\triangle_R \triangle_R ]\text{Tr}[\triangle^{\dagger}_R
\triangle^{\dagger}_R ]) \nonumber \\
&& +f_3 \text{Tr}[\triangle^{\dagger}_L \triangle_L ]\text{Tr}[\triangle^{\dagger}_R
\triangle_R ]+f_4 (\text{Tr}[\triangle_L \triangle_L ]\text{Tr}[\triangle^{\dagger}_R
\triangle^{\dagger}_R ]+\text{Tr}[\triangle_R \triangle_R ]\text{Tr}[\triangle^{\dagger}_L
\triangle^{\dagger}_L ])+\alpha_1 \text{Tr}[\Phi^{\dagger}_1 \Phi_1] \times  \nonumber \\
&& (\text{Tr}[\triangle^{\dagger}_L \triangle_L ]+ \text{Tr}[\triangle^{\dagger}_R
\triangle_R ]) +\alpha_2 ( \text{Tr}[\Phi^{\dagger}_2 \Phi_1]\text{Tr}[\triangle^{\dagger}_R
\triangle_R ]+\text{Tr}[\Phi^{\dagger}_1 \Phi_2]\text{Tr}[\triangle^{\dagger}_L
\triangle_L ]) \nonumber \\
&& + \alpha^*_2 ( \text{Tr}[\Phi^{\dagger}_1 \Phi_2]\text{Tr}[\triangle^{\dagger}_R
\triangle_R ]+\text{Tr}[\Phi^{\dagger}_2 \Phi_1]\text{Tr}[\triangle^{\dagger}_L
\triangle_L ])+\alpha_3 (\text{Tr}[\Phi_1 \Phi^{\dagger}_1\triangle_L \triangle^{\dagger}_L]
+\text{Tr}[\Phi^{\dagger}_1 \Phi_1\triangle_R \triangle^{\dagger}_R]) \nonumber \\
&& +\beta_1 (\text{Tr}[\Phi_1 \triangle_R \Phi^{\dagger}_1\triangle^{\dagger}_L]+
\text{Tr}[\Phi^{\dagger}_1 \triangle_L \Phi_1\triangle^{\dagger}_R])+\beta_2
(\text{Tr}[\Phi_2 \triangle_R \Phi^{\dagger}_1\triangle^{\dagger}_L]+
\text{Tr}[\Phi^{\dagger}_2 \triangle_L \Phi_1\triangle^{\dagger}_R]) \nonumber \\
&& +\beta_3 (\text{Tr}[\Phi_1 \triangle_R \Phi^{\dagger}_2\triangle^{\dagger}_L]+
\text{Tr}[\Phi^{\dagger}_1 \triangle_L \Phi_2\triangle^{\dagger}_R])-\mu^2_{\rho}
\rho^2+\lambda_5 \rho^4 +M \,\rho (\text{Tr}[\triangle^{\dagger}_L \triangle_L ]-
\text{Tr}[\triangle^{\dagger}_R \triangle_R ]) \nonumber \\
&& \lambda_6 \,\rho^2 (\text{Tr}[\triangle^{\dagger}_L \triangle_L ]+
\text{Tr}[\triangle^{\dagger}_R \triangle_R ])+\lambda_7 \,\rho^2
\text{Tr}[\Phi_1^{\dagger}\Phi_1]+\lambda_8 \,\rho^2 [\text{Det}[\Phi_1 ]+
\text{Det}[\Phi^{\dagger}_1] ]
\end{eqnarray*}
where $ \Phi_2 = \tau_2 \Phi^*_1 \tau_2 $. Minimizing the scalar potential we now
obtain various conditions
\begin{eqnarray}
\frac{\partial V}{\partial v_L} = \mu^2_L v_L + 2 f_1 v^3_L+f_3 v_L v^2_R +
(\beta_1 v_1v_2 +\beta_2 v^2_1 +\beta_3 v^2_2)v_R = 0
\label{eq3} \\
\frac{\partial V}{\partial v_R} = \mu^2_R v_R + 2 f_1 v^3_R+ f_3 v_R v^2_L +
(\beta_1 v_1v_2 +\beta_2 v^2_1 +\beta_3 v^2_2)v_L = 0
\label{eq4}
\end{eqnarray}
where $\mu^2_L$ and $\mu^2_R$ are effective mass terms of $ \triangle_L$ and
$\triangle_R$ given by
\begin{eqnarray*}
\mu^2_L = \mu^2_{\triangle} +M s + \lambda_6 s^2 + 2(\alpha_2+\alpha^*_2) v_1v_2+
\alpha_1 (v^2_1+v^2_2) +\alpha_3 v^2_2 \nonumber \\
\mu^2_R = \mu^2_{\triangle} -M s + \lambda_6 s^2 + 2(\alpha_2+\alpha^*_2) v_1v_2+
\alpha_1 (v^2_1+v^2_2) +\alpha_3 v^2_2
\label{mr2}
\end{eqnarray*}
Thus like in the previous case , here also the Higgs triplets $\triangle_L$
become heavier than $\triangle_R$ after the singlet $\eta$ acquires a vev at
the high scale. Equations (\ref{eq3}), (\ref{eq4}) gives
\begin{eqnarray*}
(2 M s+ (v^2_R-v^2_L)(f_3-2 f_1))v_L v_R = (v^2_L-v^2_R)(\beta_1 v_1 v_2 + \beta_2 v^2_1+\beta_3 v^2_2)
\end{eqnarray*}
Thus a nonzero vev of $\rho$ disallows those solutions for which $v_L = v_R$.
Assuming $v_L \ll v_R \ll s, M $ will give
\be
v_L = \frac{-v_R (\beta_1 v_1 v_2 + \beta_2 v^2_1+\beta_3 v^2_2)}{2 M s}
\ee
Thus we an have a small $v_L \sim  eV$ by appropriately choosing $v_R$ and $M, s$.
Here if we take $v_R$ of TeV scale then the scale of parity breaking $ M, s$ should
be low ($\sim 10^8-10^9$ GeV) so as to give $v_L \sim eV$ needed to account
for neutrino masses as we will see later.

\subsection{SUSYLR model with Higgs doublets}
\label{subsec:doublet}
We shall now study various supersymmetric left-right models.
These models are much more restrictive compared to the non-Susy models.
Although the spontaneous parity violation is one of the most important
features of the non-Susy version of the left-right symmetric models,
in the Susy left-right models with triplet Higgs scalars breaking parity
becomes very difficult and one has to extend the model to incorporate
any natural mechanism of parity violation. In this section we shall
discuss the model where the left-right symmetry is broken by Higgs
doublet scalar.

In the particle contents, the fermions belong to the fermion superfields
and we denote all the fermions and scalars by their corresponding
superfields. We can then write the particle contents of Supersymmetric Left-Right
model with Higgs doublet in terms of their superfields as
\begin{eqnarray*}
\hspace*{-2.2cm}{\rm Matter~ Superfield:} & &\, Q_L=(3,2,1,1/3), ~~Q_R =(3,1,2,1/3) \\
& & \Psi_L=(1,2,1,-1), ~~\Psi_R= (1,1,2,-1)
\end{eqnarray*}
\begin{eqnarray*}
{\rm Higgs~ Superfield:}& & \Phi_1 =(1,2,2,0), ~~\Phi_2=(1,2,2,0)   \\
& & H_L=(1,2,1,1), ~~\bar{H}_L=(1,2,1,-1), \\
& & H_R = (1,1,2,-1), ~~\bar{H}_R=(1,1,2,1), ~~\rho=(1,1,1,0)
\end{eqnarray*}
where Higgs particles with ``bar" in
the notation, helps in anomaly cancellation of the model.

The minimal Higgs doublet model without the singlet Higgs $\rho$ was disccussed in \cite{Borah:2009ra}. Here, a singlet scalar field $\rho$ is introduced, which has the special property that it is even under 
the usual parity of the Lorentz group, but it is odd under the parity that relates the gauge groups 
$SU(2)_L$ and $SU(2)_R$. This field $\rho$ is thus a scalar and not a pseudo-scalar field,
but under the D-parity transformation that interchanges $SU(2)_L$ with $SU(2)_R$, it is odd. This kind 
of work is proposed in \cite{Chang:1984qr,Hirsch:2006ft}. Although all the scalar fields are even under 
the parity of the Lorentz group, under the D-parity the Higgs sector transforms as,
\begin{eqnarray*}
H_L \leftrightarrow H_R, &\quad&
\bar{H}_L \leftrightarrow \bar{H}_R,
\nonumber \\
\Phi \leftrightarrow \Phi^\dagger, &\quad& \rho \leftrightarrow -\rho.
\end{eqnarray*}

The Higgs part of the superpotential relevant in our case is
\begin{eqnarray*}
\lefteqn{W = \mu_{ij}\text{Tr}[\tau_2 \Phi^T_i \tau_2 \Phi_j ]+
M \rho \rho +f_1 (H^T_L \Phi_i H_R +\bar{H}_L^T \Phi_i \bar{H}_R) }
\nonumber \\
&& +m_h(H^T_L\tau_2 \bar{H}_L + H^T_R \tau_2 \bar{H}_R) +
\lambda_1 \rho(H^T_L\tau_2 \bar{H}_L - H^T_R \tau_2 \bar{H}_R)
\end{eqnarray*}

The scalar potential is $V = V_F+V_D+V_{soft}$ where $V_F = \lvert F_i \rvert^2, F_i = -\frac{\partial W}{\partial \phi} $ 
is the F-term scalar potential, $V_D = D^a D^a/2, D^a = -g(\phi^*_i T^a_{ij} \phi_j)$ is the D-term 
of the scalar potential and $ V_{soft}$ is the soft 
supersymmetry breaking scalar potential. We introduce the soft Susy breaking terms to check if they alter relations between 
various mass scales in the model. The soft Susy breaking superpotential in this case is given by
\begin{eqnarray}
\lefteqn{V_{soft} = m^2_{H} H^{\dagger}_L H_L+m^2_{H} \bar{H}^{\dagger}_L \bar{H}_L+m^2_{H} H^{\dagger}_R H_R+m^2_{H} \bar{H}^{\dagger}_R \bar{H}_R+m^2_{11} \Phi^{\dagger}_1 \Phi_1} \nonumber \\
&& +m^2_{22} \Phi^{\dagger}_2 \Phi_2+ m^2_{\rho} \rho^{\dagger} \rho+(B_1H^T_L \tau_2 \bar{H}_L +B_2 H^T_R \tau_2 \bar{H}_R + B \mu_{ij}\text{Tr}[\tau_2 \Phi_i \tau_2 \Phi_j ]+h.c.) \nonumber \\
&& +(A_1 H^T_L \Phi_i H_R +A_2 \bar{H}_L \Phi_i \bar{H}_R +A_3 (\rho H^T_L \tau_2 \bar{H}_L - \rho H^T_R \tau_2 \bar{H}_R) +h.c.)
\end{eqnarray}
where all the parameters $m_H, m_{11}, m_{22}, B, A$ are of the order of Susy breaking scale $M_{susy} \sim \text{TeV}$.
We denote the vev of the neutral components of $\Phi_1, \Phi_2, H_L, \bar{H}_L, H_R, \bar{H}_R$ and $\rho$
as $\langle (\Phi_1)_{11} \rangle = v_1,~ \langle (\Phi_2)_{22} \rangle = v_2,~ \langle H_L,\bar{H}_L \rangle 
= v_L,~ \langle H_R,\bar{H}_R \rangle =v_R,~ \langle \rho \rangle = s$. 

%

Minimizing the potential with respect to $v_L, v_R$, we get the relations
\begin{eqnarray}
\lefteqn{\frac{\partial V}{\partial v_L}=-\mu^2_L (2v_L) +2v_Lv^2_R f^2_1 +f_1v_R(m_h+4 \mu)(v_1+v_2) } \nonumber \\
&& +(2m^2_H-m^2_h)v_L+A_1 s v_L+\frac{A_2 v_1 v_R}{2} + \lambda^2_1 v_L (v^2_R-v^2_L) = 0
\label{eq5}
\end{eqnarray}
\begin{eqnarray*}
\Rightarrow \frac{v_L}{v_R} = \frac{f_1(m_h+4 \mu )(v_1+v_2)+\frac{A_1 v_1}{2}}{2 \mu^2_L -2f^2_1v^2_R - \lambda^2_1 (v^2_R-v^2_L)-A_2s}
\end{eqnarray*}
\begin{eqnarray}
\lefteqn{\frac{\partial V}{\partial v_R}=-\mu^2_R (2v_R) +2v_Rv^2_L f^2_1 +f_1v_R(m_h+4 \mu)(v_1+v_2) } \nonumber \\
&& +(2m^2_H-m^2_h)v_R-A_2 s v_R+\frac{A_1 v_1 v_L}{2} -\lambda^2_1 v_L (v^2_R-v^2_L) = 0
\label{eq6}
\end{eqnarray}
where $\mu^2_L, \mu^2_R $ are given by
\begin{eqnarray*}
\mu^2_L = \frac{1}{4}[2(m_h+\lambda_1 s)^2 -4 Ms\lambda_1 -f^2_1(v^2_1 +v^2_2)]
\end{eqnarray*}
\begin{eqnarray*}
\mu^2_R = \frac{1}{4}[2(m_h-\lambda_1 s)^2 +4 Ms\lambda_1 -f^2_1(v^2_1 +v^2_2)]
\end{eqnarray*}
From equations (\ref{eq5}), (\ref{eq6}) we get
$$ (A_1v_1+4(f^2_1+\lambda^2_1)v_Lv_R+2f_1(v_1+v_2)(m_h+4\mu))(v^2_R-v^2_L)+(4sA_2+8\lambda_1s(M-m_h))v_Lv_R= 0 $$
which shows that the minimization disallows the solutions where $v_L = v_R$. Assuming
$ v_L \ll v_{1,2},\mu, A \ll s, M, m_h $ and $v_L \ll v_R$ the above expression gives rise to
\begin{equation}
v_L = \frac{v_R (2f_1m_h(v_1+v_2)+4(f^2_1+\lambda^2_1 )v_Lv_R+A_1 v_1)}{8(m_h-M)s \lambda_1 +4sA_2}
\end{equation}
Thus by appropriate choice of $m_h, M, s $ we can have TeV scale $SU(2)_R$ breaking scale $v_R$ as well
as $v_L/v_R \sim (10^{-6}-10^{-9})$ which is necessary to account for small neutrino masses as we will see later. For example, if we set
$$m_h \sim M \sim s \sim 10^{16} \, \text{GeV} \quad \text{D-parity breaking scale}$$
and allow $2m_h -M \sim 10^{8}$ GeV by appropriate fine tuning then the above relation will give rise to the desired ratio $v_L/v_R \sim 10^{-6}$. For such a choice of scales we can fine tune the parameters to get a light $H_R$ having mass $\mu_R \sim v_R \sim \text{TeV} $ and a heavy $H_L$ having mass $\mu_L \sim s, M \sim 10^{16}$ GeV. This will be important in the renormalization group running of the couplings as we will see later.
\subsection{SUSYLR model with Higgs triplets}
\label{subsec:triplet}
The particle contents of Supersymmetric Left-Right model with Higgs triplets 
in terms of their superfields are 
\begin{eqnarray*}
\hspace*{-2.2cm}{\rm Matter~ Superfield:} & &\, Q=(3,2,1,1/3), ~~Q^c =(3,1,2,1/3) \\
& & L=(1,2,1,-1), ~~L^c=(1,1,2,-1)
\end{eqnarray*}
\begin{eqnarray*}
{\rm Higgs~ Superfield:}& & \Phi_1 =(1,2,2,0), ~~\Phi_2=(1,2,2,0)   \\
& & \Delta=(1,3,1,2), ~~\bar{\Delta}=(1,3,1,-2), \\
& & \Delta^c=(1,1,3,-2), ~~\bar{\Delta}^c=(1,1,3,2), ~~\rho=(1,1,1,0)
\end{eqnarray*}
The left-right symmetry could be broken by either doublet Higgs scalars or triplet Higgs
scalar. We will show that for a minimal choice of parameters, it is convenient to break
the group with a triplet Higgs scalar. As pointed out in \cite{Aulakh:1997ba} the bidoublets are doubled to achieve
a non-vanishing Cabibbo-Kobayashi-Maskawa quark mixing and the number 
of triplets is doubled for the sake of anomaly cancellation.

The superpotential for this theory is given by
\begin{eqnarray}
W &=& \nonumber Y^{(i)_{q}} Q^{T} \tau_{2} \Phi_{i} \tau_{2} Q^{c} 
  + Y^{(i)_{l}} L^{T} \tau_{2} \Phi_{i} \tau_{2} L^{c}\\ 
\nonumber 
&& + ~i (f L^{T} \tau_{2} \Delta L +f^{*} L{^{c}}^{T} \tau_{2} 
  \Delta^{c} L^{c} ) +M \rho^2\\
&& + ~m_{\Delta} \textrm{Tr}(\Delta \bar{\Delta}) + m^{*}_{\Delta} 
  \textrm{Tr}(\Delta^{c} \bar{\Delta}^{c})+ \mu_{ij} 
  \textrm{Tr}(\tau_{2}\Phi^{T}_{i} \tau_{2} \Phi_{j}).
\end{eqnarray}
All couplings $Y^{(i)_{q,l}}$, $\mu_{ij}$, $\mu_{\Delta}$, $f$ in the
above potential, are complex with the the additional constraint that 
$\mu_{ij}$, $f$ and $f^{*}$ are symmetric matrices.
The scalar potential is $V = V_F+V_D+V_{soft}$ where $V_F = \lvert F_i \rvert^2, F_i = 
-\frac{\partial W}{\partial \phi} $ is the F-term scalar potential, $V_D = D^a D^a/2, 
D^a = -g(\phi^*_i T^a_{ij} \phi_j)$ is the D-term 
of the scalar potential and $ V_{soft}$ is the soft supersymmetry breaking terms in the 
scalar potential. In the particular model, the soft-susy breaking terms are given by
\begin{eqnarray}
\lefteqn{V_{soft} = m^2_{\delta} Tr[\Delta^{\dagger} \triangle ]+m^2_{\delta} Tr[\bar{\Delta}^{\dagger} 
\bar{\Delta} ]+m^2_{\delta} Tr[(\Delta^c)^{\dagger} \Delta^c ]+m^2_{\delta} Tr[\bar{\Delta^c}^{\dagger} 
\bar{\Delta^c} ]+m^2_{11} \Phi^{\dagger}_1 \Phi_1} \nonumber \\
&& +m^2_{22} \Phi^{\dagger}_2 \Phi_2+ m^2_{\rho} \rho^{\dagger}\rho+(B \mu_{ij}\text{Tr}[\tau_2 \Phi_i \tau_2 \Phi_j ]+ 
A\rho(\text{Tr}[\Delta \bar{\Delta}]-\text{Tr}[\Delta^c \bar{\Delta}^c]+h.c.)
\label{vsoft2}
\end{eqnarray}
where all the parameters in the soft supersymmetry breaking scalar potential is of the 
order of supersymmetry breaking scale $M_{susy} \sim \text{TeV}$.
We denote the vev of the neutral components of $\Phi_1, \Phi_2, \Delta, \bar{\Delta}, \Delta^c,
\bar{\Delta}^c$ and $\rho$ as
$$ \langle (\Phi_{1})_{11} \rangle = v_1, \quad \langle (\Phi_2)_{22} \rangle = v_2, \quad \langle \Delta, 
\bar{\Delta} \rangle = v_L,\quad \langle \Delta^c,\bar{\Delta}^c \rangle =v_R, \quad  \langle \rho \rangle = s$$
Minimizing the scalar potential with respect
to $v_L, v_R$ we get
\be
\frac{\partial V}{\partial v_L} = v_L[2(m_{\Delta}+\lambda_1 s)^2+2\lambda^2_1
(v^2_L-v^2_R)+A s+2m^2_{\delta}] = 0 \nonumber \\
\ee
\be
\Rightarrow v^2_R-v^2_L = \frac{2m^2_{\delta}+(A+2\lambda_1 M)s + 2(m_{\Delta}+\lambda_1s)^2}{2\lambda^2_1}
\label{eq7}
\ee
\be
\frac{\partial V}{\partial v_R} = v_R[2(m_{\Delta}-\lambda_1 s)^2-2\lambda^2_1
(v^2_L-v^2_R)-A s+2m^2_{\delta}] = 0 \nonumber \\
\ee
\be
\Rightarrow v^2_R-v^2_L = \frac{-2m^2_{\delta}+(A+2\lambda_1 M)s -2(m_{\Delta}-\lambda_1s)^2}{2\lambda^2_1}
\label{eq8}
\ee
Also
\be
v_R\frac{\partial V}{\partial v_L}-v_L\frac{\partial V}{\partial v_R}=4 v_L v_R
[2(Ms+2 m_{\Delta}s)\lambda_1+2\lambda^2_1 (v^2_L-v^2_R)+As] = 0 \nonumber \\
\ee
\be
\Rightarrow v^2_R-v^2_L = \frac{2\lambda_1(Ms+2 m_{\Delta})+As}{2\lambda^2_1}
\label{eq9}
\ee
Thus the minimization conditions disallows solutions with $v_L = v_R$. But from equations
(\ref{eq7}), (\ref{eq8}), (\ref{eq9}) it can be seen that it is difficult to adjust the
various scales $M , s, m_{\triangle} $ so as to satisfy them simultaneously and giving
rise to a TeV scale $v_R$ and an eV scale $v_L$. Thus we need to add more particles to
the above particle content which can give rise to spontaneous D-parity breaking with a
TeV scale $v_R$. This scenario of minimal SUSYLR model with parity odd singlet was studied long ago and was
shown \cite{Kuchimanchi:1993jg} that the charge-breaking vacua have a
lower potential than the charge-preserving vacua and as such the ground
state does not conserve electric charge
\subsection{SUSYLR model with Higgs triplets and bitriplet}
The minimal left-right supersymmetric models with triplet Higgs bosons leads to several 
nettlesome obstructions which may be considered to be a guidance towards a unique consistent 
theory. One of the most important problems is the spontaneous breaking of left-right symmetry 
and substantial amount of work has been done to cure this problem. This can be 
cured either by adding some extra fields to the minimal particle content \cite{Kuchimanchi:1993jg} 
or with the help of non-renormalization operator \cite{Aulakh:1998nn}. There is another solution 
to the problem, which resembles the non-supersymmetric solution, relating the vacuum expectation 
values (vevs) of the left-handed and right-handed triplet Higgs scalars to the Higgs bi-doublet 
vev through a seesaw relation. The novel feature consists in the introduction of a bitriplet Higgs 
and another Higgs singlet under left-right group \cite{Patra:2009wc}. We will try to extremize the 
full potential of this particular model and see what are the mass scales, different vevs coming out 
from the extremization.

We now present our model, where we include a bi-triplet $\eta=(1,3,3,0)$ and a parity odd singlet field $\rho=(1,1,1,0)$ in the minimal supersymmetric left-right symmetric model with triplet Higgs discussed in 
previous subsection. These fields are vector-like and hence do not contribute to anomaly, so 
we consider only one of these fields. Under parity, these fields transform as $\eta \leftrightarrow \eta$ and 
$\rho \leftrightarrow -\rho$. The superpotential for the model is
written in the more general tensorial notation \cite{Patra:2009wc} as follows
\begin{eqnarray}
\lefteqn{W = f \eta_{\alpha i} \Delta_{\alpha} \Delta^c_i+f^* \eta_{\alpha i}
\bar{\Delta}_{\alpha}\bar{\Delta}^c_i+ \lambda_1 \eta_{\alpha i} \Phi_{am}
\Phi_{bn} (\tau^{\alpha} \epsilon)_{ab} (\tau^{i} \epsilon)_{mn}+m_{\eta}
\eta_{\alpha i}\eta_{\alpha i}} \nonumber \\
&& +M_{\Delta}(\Delta_{\alpha}\bar{\Delta}_{\alpha}+\Delta^c_i
\bar{\Delta}^c_i) +\mu \epsilon_{ab}\Phi_{bm}\epsilon_{mn}\Phi_{an}+ m_{\rho}
\rho^2 +\lambda_2 \rho (\Delta_{\alpha}\bar{\Delta}_{\alpha}-\Delta^c_i
\bar{\Delta}^c_i)
\end{eqnarray}
where $\alpha, a, b$ are $SU(2)_L$ and $i, m, n$ are $SU(2)_R$  indices. The symmetry breaking pattern in this model is
$$ SU(2)_L\times SU(2)_R\times U(1)_{B-L} \times P \quad \underrightarrow{\langle \rho \rangle} \quad SU(2)_L\times SU(2)_R\times U(1)_{B-L} $$
$$ \underrightarrow{\langle \bigtriangleup_c \rangle} \quad SU(2)_L\times U(1)_Y \quad \underrightarrow{\langle \Phi \rangle} \quad U(1)_{em} $$
Denoting the
vev's as $\langle \Delta_- \rangle = \langle \bar{\Delta}_+ \rangle = v_L,~
\langle \Delta^c_+ \rangle = \langle \bar{\Delta}^c_- \rangle = v_R, ~\langle
\Phi_{+-} \rangle = v, ~\langle \Phi_{-+} \rangle = v', ~\langle \eta_{+-} \rangle =
u_1, ~\langle \eta_{-+} \rangle = u_2, ~\langle \eta_{00} \rangle = u_0$ and $\langle \rho \rangle=s $.

The scalar potential is $V = V_F+V_D+V_{soft}$ where $V_F = \lvert F_i \rvert^2, F_i = 
-\frac{\partial W}{\partial \phi} $ is the F-term scalar potential, $V_D = D^a D^a/2, D^a = 
-g(\phi^*_i T^a_{ij} \phi_j)$ is the D-term 
of the scalar potential and $ V_{soft}$ is the soft supersymmetry breaking terms in the 
scalar potential. In the particular model, the soft-susy breaking terms are given by
\begin{eqnarray}
\lefteqn{V_{soft}=V_{soft}(\text{containing $\triangle$ and $\Phi$}) +m_{\eta (soft)}\eta^{\dagger}_{\alpha i} 
\eta_{\alpha i}} \nonumber \\
&& +(A_2 \eta_{\alpha i} \Phi_{am}\Phi_{bn} (\tau^{\alpha} \epsilon )_{ab} (\tau^{i} \epsilon )_{mn}+A_3 
(\eta_{\alpha i} \Delta_{\alpha} \Delta^c_i ) +h.c.)
\end{eqnarray}
where $V_{soft}(\text{containing $\triangle$ and $\Phi$})$ is given by the eqn: (\ref{vsoft2}) 
in the subsection [\ref{subsec:triplet}].

Minimizing the scalar potential with respect to $v_L, v_R$ we get
\begin{eqnarray}
\lefteqn{\frac{\partial V}{\partial v_L} = \mu^2_L (2v_L) +2 \lambda^2_2 v_L(v^2_L-v^2_R)
+2(fu_1+f^* u_2)M_{\Delta} v_R+v_R(f+f^*)[2m_{\eta} (u_1 } \nonumber \\
&& + u_2+u_3)+\lambda_1 v^2 +v_Lv_R(f+f^*)]+4v_Lm^2_{\delta}+2Av_L s+A_3 v_R(u_1+u_2+u_3) = 0
\label{eq10}
\end{eqnarray}
\begin{eqnarray}
\lefteqn{\frac{\partial V}{\partial v_R} = \mu^2_R (2v_R) -2 \lambda^2_2 v_R(v^2_L-v^2_R)
+2(fu_1+f^* u_2)M_{\Delta} v_L+v_L(f+f^*)[2m_{\eta} (u_1} \nonumber \\
&& + u_2+u_3)+\lambda_1 v^2 +v_Lv_R(f+f^*)]+4v_Rm^2_{\delta}-2Av_Rs+A_3 v_L(u_1+u_2+u_3) = 0
\label{eq11}
\end{eqnarray}
Where the effective mass terms $\mu^2_L, \mu^2_R $ are given by
\be
\mu^2_L = (M_{\Delta}+\lambda_2 s)^2 + \lambda_2 m_{\rho}s +\frac{1}{2}(f^2u^2_1+f^{*2}u^2_2)
\ee
\be
\mu^2_R = (M_{\Delta}-\lambda_2 s)^2 -\lambda_2 m_{\rho}s +\frac{1}{2}(f^2u^2_1+f^{*2}u^2_2)
\ee
Thus after the singlet field $\rho$ acquires a vev the degeneracy of the Higgs triplets goes
away and the left handed triplets being very heavy get decoupled whereas the right handed
triplets can be as light as 1 TeV by appropriate fine tuning in the above two expressions.
Assuming $v_L \ll v,v',\mu,A \ll m_{\rho},s$  and $v_L \ll v_R$ we get from equations (\ref{eq10}), (\ref{eq11}):
\be
v_L = \frac{-v_R[M_{\Delta} u_2 f^* +m_{\eta} (u_2+u_3)(f+f^*)+u_1(fM_{\Delta}+
m_{\eta}(f+f^*)]}{2m_{\rho}s\lambda_2 +4 M_{\Delta}s\lambda_2+2As}
\ee
Thus we can get a small $v_L (\sim \text{eV}) $ and a TeV scale $v_R$ by appropriate choice
of $M_{\Delta}, m_{\eta}, m_{\rho}, s$. We take the vev of the bitriplet $u \ll M_Z$. Thus 
if we want $v_R \sim 1$ TeV then the above relation will give us an eV scale $v_L$ only if 
the scale of parity breaking is kept low that is, $s \sim m_{\rho} \sim M_{\Delta} 
\sim 10^{10}$ GeV. Thus in such a type II seesaw dominated case, the right handed triplets 
$\Delta^c$ will be as light as $\mu_R \sim v_R \sim 1$ TeV and the left handed triplets 
$\Delta$ as heavy as $\mu_L \sim 10^{10}$ GeV by appropriate fine tuning of the parameters. 
However as we will see later, such a light Higgs triplet with $B-L$ charge 2 spoils the gauge 
coupling unification. Hence we are forced to keep the intermediate symmetry breaking scale 
$M_R$ close to the unification scale.
\section{Gauge Coupling Unification}
\label{sec:gauge}
Grand unified theories (GUTs) offer the possibility of unifying the
three gauge groups viz., $SU(3)$, $SU(2)$ and $U(1)$ of the standard
model into one large group at a high energy scale $M_U$. This scale is
determined as the intersection point of the $SU(3)$, $SU(2)$ and $U(1)$
couplings. The particle content of the theory completely determines the
variation of the couplings with energy. It is hard to achieve low intermediate 
scale without taking into account the effect of D-parity breaking in the renormalization groupl equations (RGEs).
We have seen in the previous section that in spontaneous D-parity breaking models,
the minimization of the scalar potential simultaneously allows us to have right handed
scale $v_R$ of the order of TeV and tiny neutrino masses from seesaw mechanisms.
However the evolution of gauge couplings will be very different in models with Higgs
triplets and with Higgs doublets. In this section we study the renormalization group
evolution of the gauge couplings and see if unification at a high scale ($\sim 10^{16}$ GeV) 
allows us to have a TeV scale $v_R$. Similar analysis were done in
\cite{Dev:2009aw,Patra:2010ks} for Higgs doublet case. Here we use the $U(1)$ normalization
constant $\sqrt{\frac{3}{8}}$ as in \cite{Setzer:2005hg}. We restrict our study to the supersymmetric case only. The gauge coupling unification in the non-supersymmetric versions of such models were studied before and can be found in \cite{Chang:1984qr,Bhatt:2008dg}.
\subsection{Unification in SUSYLR model with Higgs doublets}
\label{subsec:rgdoublet}
We will study the evolution of couplings according to their respective beta functions 
with the account of spontaneous D-parity breaking. The renormalization group equations(RGEs) 
for this model cane be written as
\begin{equation}
\frac{d \alpha_{i}}{d t}=\alpha^{2}_{i} [b_{i}+\alpha_{j} b_{ij}+ O(\alpha^{2}) ]
\end{equation}
where, $t=2\pi\, ln(M)$ (M is the varying energy scale), $\alpha_i=\frac{g_i^2}{4 \pi}$ is the 
coupling strength. Also $b_i$ and $b_{ij}$ are the one loop and two loop beta coefficients and 
we will study only the one loop contributions to RGEs \cite{Setzer:2005hg}. The indices $i,j=1,2,3$ refer to the gauge
group $U(1)$, $SU(2)$ and $SU(3)$ respectively.

The particle content of SUSYLR model with Higgs doublets is shown in subsection 
[\ref{subsec:doublet}]. It turns out that the minimal particle content is not enough 
for proper gauge coupling unification. For required unification purposes we add two 
copies of $\delta (1,1,1,2), \bar{\delta} (1, 1,1,-2)$ at the $SU(2)_R$ breaking scale. 
The beta functions are given as
\begin{itemize}
\item Below the Susy breaking scale $M_{susy}$ the beta functions are same as those of the
standard model
$$ b_s = -7, \quad b_{2L} = -\frac{19}{6} \quad  b_Y = \frac{41}{10}$$
\item For $ M_{susy} < M < M_R $ , the beta functions are same as those of the MSSM
$$ b_s = -9+2n_g,\quad b_{2L} = -6+2n_g+\frac{n_b}{2}, \quad  b_Y = 2n_g+\frac{3}{10}n_b $$
\item For $ M_R < M < \langle \rho \rangle $ the beta functions are
$$ b_s = -9+2n_g,\quad b_{2L} = -6 +2n_g +n_b $$
$$ b_{2R} = -6+2n_g + n_b + \frac{n_{HR}}{2},\quad b_{B-L} = 2n_g +3 n_{\delta}+ \frac{3}{4}n_{HR} $$
\item For $ \langle \rho \rangle < M < M_{GUT} $ the beta functions are
$$ b_s = -9+2n_g,\quad b_{2L} = -6 +2n_g + n_b+\frac{n_{HL}}{2}  $$
$$ b_{2R} = -6+2n_g + n_b + \frac{n_{HR}}{2},\quad b_{B-L} = 2n_g +3 n_{\delta}+ \frac{3}{4}(n_{HL}+n_{HR}) $$
\end{itemize}
\begin{figure}[t]
\centering
\includegraphics{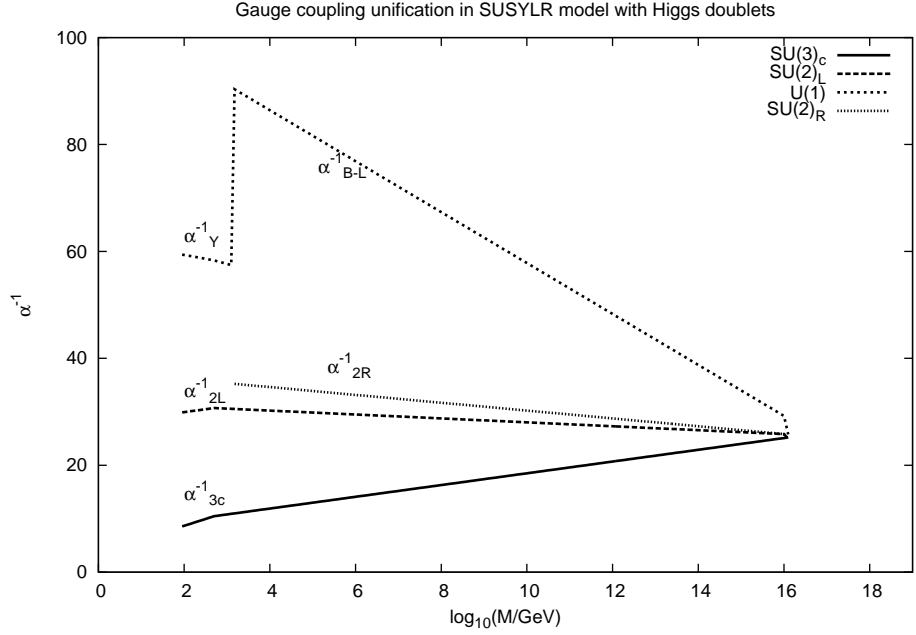}
\caption{Gauge coupling unification in SUSYLR model with Higgs doublets and $M_{susy} = 500$ GeV, $M_R = 1.5$ TeV, $M_{\rho} = 10^{16}$ GeV}
\label{fig1}
\end{figure}
where $n_g$ is the number of fermion generations and number of Higgs bidoublets $ n_b = 2$, number of Higgs 
doublets $n_{HL}=n_{HR} = 2$, number of extra Higgs singlets $n_{\delta}
= 2$. The experimental initial values for the couplings at electroweak scale $M=M_Z$ \cite{Amsler:2008zzb} are
\begin{equation}
\left(\begin{array}{cc}
\ \alpha_s (M_Z) \\
\ \alpha_{2L}(M_Z) \\
\ \alpha_{1Y}(M_Z)
\end{array}\right)
= \left(\begin{array}{cc}
\ 0.118 \pm 0.003 \\
\ 0.033493^{+0.000042}_{-0.000038} \\
\ 0.016829 \pm 0.000017
\end{array}\right)
\label{par1}
\end{equation}
The normalization condition at $M = M_{R}$ where the $U(1)_Y$ gauge coupling
merge with $SU(2)_R \times U(1)_{B-L} $ is $\alpha^{-1}_{B-L}=\frac{5}{2}
\alpha^{-1}_Y-\frac{3}{2}\alpha^{-1}_L$. Using all these we arrive at the
gauge coupling unification as shown in (\ref{fig1}). Here we have taken
$M_{susy} = 500 $ GeV, $M_R = 1.5$ TeV, $M_{\rho} = 10^{16}$ GeV.
The couplings seems to unify at a scale slightly above the D-parity breaking scale.
Thus the D-parity breaking scale need not be the same as the GUT scale, but can be
lower also. However if we make the D-parity breaking scale arbitrarily lower, the
unification wont be possible as can be seen from the figure (\ref{fig1}). Since both
the left handed and right handed Higgs doublets will contribute to the $U(1)_{B-L}$
couplings  after the D-parity breaking scale, the $\alpha^{-1}_{BL} $ will come down
sharply and meet the other couplings at some energy below the expected GUT scale.
\subsection{Unification in SUSYLR model with Higgs triplets}
\label{subsec:rgtriplet}
\begin{figure}[t]
\centering
\includegraphics{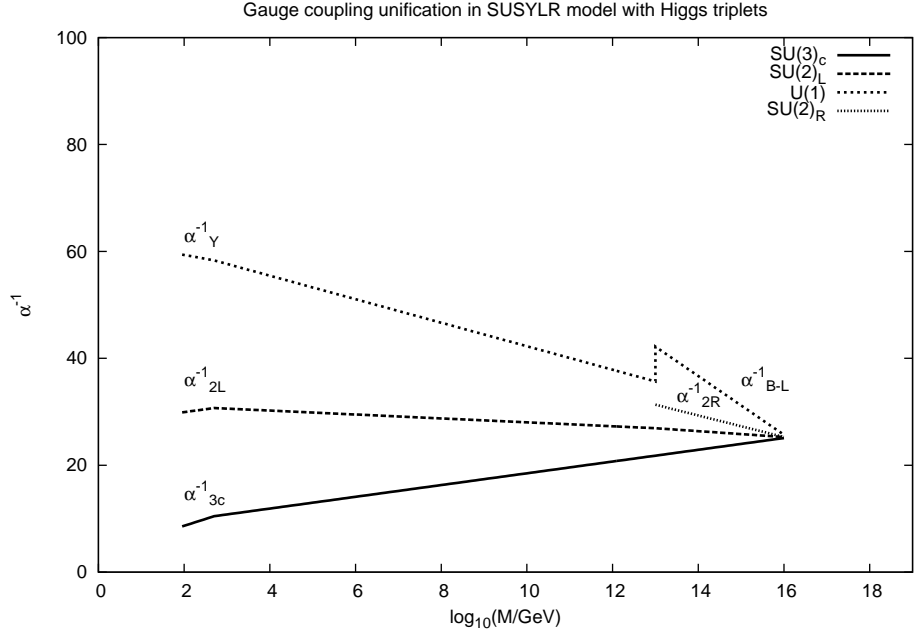}
\caption{Gauge coupling unification in SUSYLR model with Higgs triplets and  $M_R = 10^{13} $ GeV, $M_{\rho} = 10^{16}$ GeV}
\label{fig2}
\end{figure}

The particle content of SUSYLR model with Higgs triplets is shown in subsection 
[\ref{subsec:triplet}]. It is very difficult to achieve unification with low $M_R$ 
with the minimal particle content. We add a parity odd singlet $\rho(1,1,1,0)$ to achieve 
spontaneous D-parity breaking. This may change the scale of $M_R$, but it is found that the 
$M_R$ remains higher than $10^{10}$ GeV.
For unification purposes, we need in the recent model, one heavy bidoublet $\chi(1,2,2,0)$ 
has been added which gets mass at the $SU(2)_R$ breaking scale. Below the $SU(2)_R$ breaking 
scale the beta functions are similar to the MSSM as written above. The beta functions above 
this scale are
\begin{itemize}
\item For $M_R < M < M_{\rho}$ the beta functions are
$$ b_s = -9+2n_g,\quad b_{2L} = -6 +2n_g +n_b + \frac{n_{\chi}}{2}  $$
$$ b_{2R} = -6+2n_g +n_b+2n_{\triangle} + \frac{n_{\chi}}{2},\quad b_{B-L} = 2n_g +\frac{9}{2}n_{\triangle} $$
\item For $ \langle \rho \rangle < M < M_{GUT} $ the beta functions are
$$ b_s = -9+2n_g,\quad b_{2L} = -6 +2n_g +n_b+2 n_{\triangle}+\frac{n_{\chi}}{2} $$
$$ b_{2R} = -6+2n_g + n_b + 2 n_{\triangle}+\frac{n_{\chi}}{2},\quad b_{B-L} = 2n_g + 9n_{\triangle} $$
\end{itemize}
where number of Higgs triplets $n_{\triangle} = 2$, number of additional Higgs field added for unification $n_{\chi}=1,$, number of generations $ n_g = 3$,  and number of Higgs bidoublets $n_b = 2$.
Using the same initial values and normalization relations like before we arrive at the
gauge coupling unification as shown in (\ref{fig2}).
Here the unification scale $M_{GUT}$ coincides with the D-parity breaking
scale $M_{\rho}$. Lower values of $M_R$ will make the unification worse
because of the large contributions of triplets to the $U(1)_{B-L}$ beta
functions compared to the doublets in the previous case. Thus in the minimal
triplet case, both the minimization conditions as well as unification
disallow a TeV scale $v_R$. Although after adding a bitriplet, the minimization
conditions allow a TeV scale $v_R$, it wont make the unification better as
we discuss in the next subsection.
\subsection{Unification in SUSYLR model with Higgs triplets and bitriplet}
\label{subsec:bitriplet}
As we saw before, the minimization of the scalar potential in a SUSYLR model
with Higgs triplets with spontaneous D-parity breaking does not allow a TeV
scale $M_R$. The same thing is true from gauge coupling unification point of
view as shown in the previous subsection. Now we consider the SUSYLR model with
Higgs triplet as well a bitriplet \cite{Patra:2009wc}. For unification purposes
we include two pairs of heavy colored superfields $\chi (3, 1,1,0), \bar{\chi}(\bar{3},1,1,0) $ which decouple after
the $SU(2)_R $ breaking scale $M_R$. The beta functions above $ M_R$ are

\begin{figure}[htb]
\centering
\includegraphics{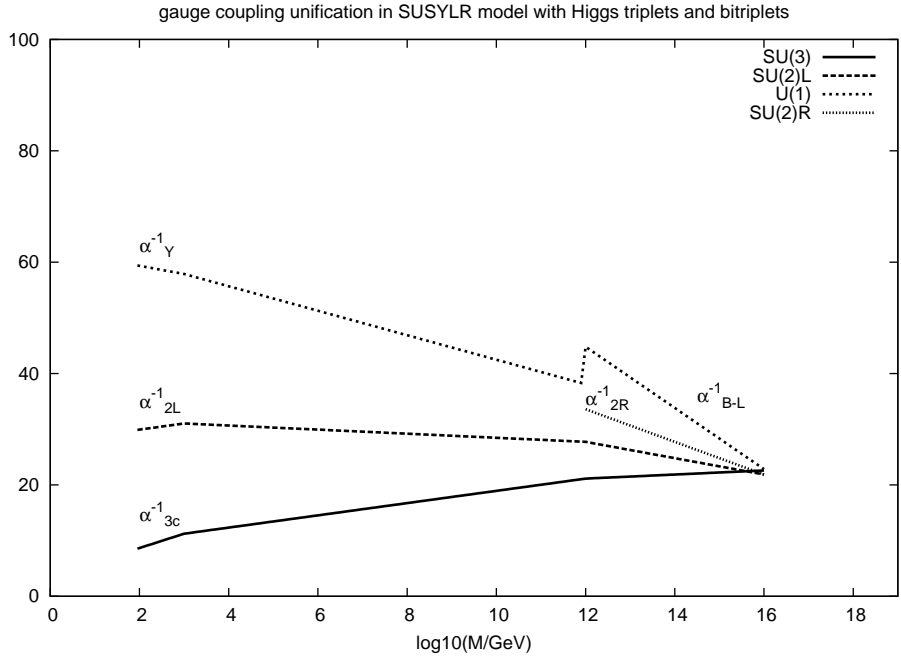}
\caption{Gauge coupling unification in the bitriplet model with two extra pairs of colored 
superfields $\chi(3,1,1,0), \bar{\chi}(\bar{3},1,1,0)$, $M_{susy} = 1$ TeV, $M_R = 10^{12} $ GeV, $M_{\rho} = 10^{16}$ GeV. The extra superfields decouple below the scale $M_R$.}
\label{fig3}
\end{figure}

\begin{itemize}
\item For $M_R < M < M_{\rho}$ the beta functions are
$$ b_s = -9+2n_g + \frac{n_{\chi}}{2},\quad b_{2L} = -6 +2n_g +n_b + 2n_{\eta} $$
$$ b_{2R} = -6+2n_g + n_b+2n_{\triangle} + 2n_{\eta},\quad b_{B-L} = 2n_g +\frac{9}{2}n_{\triangle}$$
\item For $ \langle \rho \rangle < M < M_{GUT} $ the beta functions are
$$ b_s = -9+2n_g+  \frac{n_{\chi}}{2},\quad b_{2L} = -6 +2n_g +n_b+2 n_{\triangle}+2 n_{\eta} $$
$$ b_{2R} = -6+2n_g + n_b + 2 n_{\triangle}+ 2 n_{\eta},\quad b_{B-L} = 2n_g + 9n_{\triangle} $$
\end{itemize}

where number of Higgs triplets $n_{\triangle} = 2$, number of colored Higgs $n_{\chi}=3$, number of generations $n_g = 3$, number of Higgs bidoublets  $n_b = 2$ and number of Higgs bitriplets $n_{\eta}=1 $.
Using the same initial values and normalization relations like before
we arrive at the gauge coupling unification as shown in (\ref{fig3}).
Here the unification scale is the same as the D-parity breaking scale.
Similar to the case with just Higgs triplets, here also lower value of
$M_R$ makes the unification look worse. Thus although minimization of
the scalar potential allows the possibility of a TeV scale $M_R$ in
this model, the gauge coupling unification criteria rules out such a
possibility.
\section{Neutrino mass in SUSYLR model with Higgs doublets}
\label{subsec:neumass}
In left-right symmetric models with only doublet scalar fields, the question of neutrino masses 
has been discussed in details. We shall try to restrict ourselves as close as possible
to these existing non-supersymmetric models, and check the consistency of these solutions
when D-parity is broken spontaneously in the present SUSYLR model. 

 We introduced a singlet fermionic superfield $S$ to the particle content of the model discussed in subsection 
[\ref{subsec:doublet}]. This kind of model has been discussed without the D-parity breaking effect and from the 
neutrino mass prospective \cite{Borah:2009ra}. The effect of this singlet field has been accounted in the RGEs shown in subsection 
[\ref{subsec:rgdoublet}]. With the addition of this singlet fermion, the superpotential and resulting neutrino 
mass matrix become 
\begin{equation}
W = {\cal M}_{ij} S_{i} S_{j}+F_{ij} {\Psi_{L}}_{i} S_{j} H_{L}
+ F'_{ij} {\Psi_{R}}_{i} S_{j} H_{R},
\label{eq:spot}
\end{equation}
and 
\begin{equation}
W_\textrm{neut} = (\nu_i \quad  N^c_i \quad  S_i)
        \left( \begin{array}{ccc}
                0        & (M_N)_{ij}   & F_{ij} v_{L} \\
              (M_N)_{ji} &    0         & F'_{ij} v_{R} \\
              F_{ji} v_{L} & F'_{ji} v_{R} & {\cal M}_{ij}
        \end{array} \right)
        \left( \begin{array}{c}
            \nu_j \\ N^c_j \\ S_j
        \end{array} \right).
\label{eq:spotmatrix}
\end{equation}
where $M_N$ is the general Dirac term coming from the term $(M_N)_{ij} \nu_i N^c_j$.
In the above mass matrix, the mass of the singlet $\mathcal{M}_{ij}$ and
the vev of the right-handed Higgs doublet $v_R$ are heavy, while $M_N$ and
vev of the left-handed Higgs doublet $v_L$ are of low scale.

The resulting light neutrino mass matrix after diagonalizing the above mass matrix 
is 
\begin{eqnarray}
M_{\nu} &=& - M_N M_R^{-1} M_N^T - (M_N H + H^T M_N^T)
    \left(\frac{v_{L}}{v_{R}}\right),
\label{eq:typeIII}
\\
\textrm{where,}\qquad
H &\equiv& \left( F' \cdot F^{-1}\right)^{T},
\label{eq:Hmatrix}
\\
M_{R} &=& (F \,v_{R}) {\cal M}^{-1} (F^T v_{R}).
\label{eq:calM}
\end{eqnarray}
Here we can see that the first term in eqn (\ref{eq:typeIII}) is the usual type-I seesaw contribution 
and the second term is an another seesaw term giving rise to a double seesaw mechanism. This second term contribution to $\nu$ mass 
will dominate over type-I if the elements of the matrix ${\cal M}_{ij}$ are small compared to 
the contribution of $H$ term. It is clear from the eqn (\ref{eq:calM}) that the scale of $M_R$ found to 
be TeV for ${\cal M}_{ij}=$1 TeV, $v_R=$1 TeV which is automatically comes from the minimization of the 
potential and consistent with the renormalization-group evolutions which has already studied in subsection [\ref{subsec:rgdoublet}] 
and $F$ of the order of unity. With the mass scales and $M_N$ of the order of MeV, we can found neutrino mass to 
be eV.

{\bf \hspace{-0.6cm} Neutrino mass in case of Fermionic triplet:}

Let us introduce fermionic triplets (one for each family) order to realize the
type III seesaw mechanism:
\begin{equation*}
\Sigma_L = \frac{1}{2} \left(
\begin{array} {cc}
 \Sigma_L^0  &  \sqrt{2} \Sigma^+_L \\
 \sqrt{2} \Sigma^-_L  & - \Sigma_L^0
\end{array}
\right) \ \equiv \ (3,1,1,0),
\end{equation*}
and
\begin{equation*}
\Sigma_R = \frac{1}{2} \left(
\begin{array} {cc}
 \Sigma_R^0  &  \sqrt{2} \Sigma^+_R \\
 \sqrt{2} \Sigma^-_R  & - \Sigma_R^0
\end{array}
\right) \ \equiv \ (1,3,1,0),
\end{equation*}
Under left-right parity transformation
one has the following relations
\begin{equation*}
\Sigma_L \longleftrightarrow \Sigma_R.
\end{equation*}

In the context of lepton masses, the relevant term in the Lagrangian is
$${\cal L}_\ell=\bar{\ell}_L (Y_1 \Phi +Y_2 \tilde{\Phi}) \ell_R+h.c. $$
where $\tilde{\Phi}=\tau_2 \Phi \tau_2$. Once the bidoublet $\Phi$ takes vev. i.e
$v_1=\langle \phi^0_1 \rangle$ and $v_2=\langle \phi^0_2 \rangle$, the Dirac mass matrix
for the neutrinos is
$$m^D_\nu=Y_1 v_1 +Y_2 v_2 $$
The relevant Yukawa terms that gives masses (for the type III seesaw mass matrix) to the
three generations of leptons are given by
\begin{eqnarray}
{\cal L}^{III}_\nu &=& \ h_{ij} \ell_{iL}^T \ C \ i \sigma_2 \ \Sigma_{jL} \ H_L
\ + g_{ij}\ \ell_{iR}^T \ C \ i \sigma_2 \ \Sigma_{jR} \ H_R  \nonumber \\
&+& \ M_\Sigma \ Tr \left( \Sigma_L^T \ C \ \Sigma_L \ + \ \Sigma_R^T \ C \Sigma_R \right) +
 h.c.
\end{eqnarray}
Once the Higgs doublets gets vev i.e,$v_L=\langle H_L^0 \rangle$ and $v_R=\langle H_R^0 \rangle$,
$SU(2)_L \otimes SU(2)_R$ is broken spontaneously. Now the mass matrix in the basis
$\left(\nu_L, \ \nu_R, \ \Sigma^0_R, \ \Sigma^0_L \right)$ reads as:
\begin{equation}
M_{\nu}^{III} = \left(
\begin{array} {cccc}
 0 & m_\nu^D & 0 & h v_L \\
(m_\nu^D)^T & 0 & g v_R & 0
\\
0 & g^T v_R & M_{\Sigma} & 0 \\
h^T v_L & 0 & 0 & M_{\Sigma}
\end{array}
\right).
\end{equation}
As one expects the neutrino masses
are generated through the Type I + Type III seesaw mechanisms
and one has a $double$ seesaw mechanism since the mass of
the right-handed neutrinos are generated through
the Type III seesaw once we integrate out $\Sigma_R^0$.

The neutrino mass formula derived from the above mass matrix is given by
\begin{equation}
m_{\nu_L} = \frac{1}{v_R^2\,(g^T g)}\,[m_{\nu}^D \,M_{\Sigma}\, (m_{\nu}^D)^T-v_R \,
v_L\, m_{\nu}^D \,(g\,h)^T-v_R\, v_L\, (g\,h)\,(m_{\nu}^D)^T]
\label{nu1}
\end{equation}
with right handed neutrino masses
\begin{equation}
\label{eqn:mr}
M_{R} = v_R^2 \,g \, \left( M_{\Sigma} \right)^{-1}\, g^T.
\end{equation}

We take the Dirac mass of the all the three neutrinos to be of MeV order. This fixes the scale of the $M_\Sigma$ and $M_R$ so as to give rise to eV 
scale neutrino masses on the left hand side of above relation [\ref{nu1}]. If we assume that the first 
term of [\ref{nu1}] will dominate then the seesaw relations will become
$m_{\nu}=\frac{m^2_e}{M_{R}}$. As $m_e=0.5$ MeV, we need the values of the right handed Majorana neutrino
as: $M_{R}=10^{3}$ GeV to
have 0.1 eV light neutrino mass. We can arrive at the appropriate value of $M_R$ by choosing $g$ and 
$M_{\Sigma}$. Since we are taking $v_R \sim 1 \text{TeV}$ hence to get $M_R \ge 1 \text{TeV}$ we must 
have $M_{\Sigma} \le 1 \text{TeV}$. Once the scale of right handed Majorana neutrino gets
fixed by the light neutrino mass, we can find the values of $M_\Sigma$ and $v_R$.
We have taken the Yukawa couplings as $g,h < 1$, $v_R =10^{3}$ GeV in Eq. [\ref{eqn:mr}]
and these lead to triplet fermion masses :$M_{\Sigma}\sim 10^3$ GeV. \\
\indent If $M_{\Sigma} << 1 \text{TeV}$ and $v_R \sim 1 \text{TeV}$, then the first term of the above 
neutrino mass formula becomes to small to give rise to neutrino masses. In that case the second and 
the third term in the equation [\ref{nu1}] can contribute to the neutrino masses if $v_L/v_R \sim 10^{-6}$. 
And such a ratio can naturally be achieved (even if we have a TeV scale $v_R$) by choosing various symmetry 
breaking scales and mass parameters as we discussed in section [\ref{sec2}].

{\bf \hspace{-0.6cm}Role of $\Sigma_L, \Sigma_R$ in unification:}

The fermion triplets with $U(1)_{B-L}$ charge zero contributes to the $SU(2)_L$ and 
$SU(2)_R$ gauge coupling running. As discussed above, for the seesaw purposes we have to take low 
values of $M_{\Sigma} <= v_R$ which will ruin the gauge coupling unification for a TeV scale $SU(2)_R$ 
breaking scale $v_R$. Unification and small neutrino mass are possible only if $SU(2)_R$ breaking 
scale as well as mass of the triplet fermions are close to the unification scale. However if we add 
fermion singlet in place of triplets then there is no constraints from unification point of view on 
$v_R$ and $M_{\Sigma}$. The mass matrix becomes $3 \times 3 $ in this case. Thus in Supersymmetric 
left-right model with Higgs doublets, we can achieve unification with TeV scale $SU(2)_R$ breaking 
scale only if fermion singlet is added in place of triplets as in the conventional type III seesaw.
\section{Neutrino mass in SUSYLR model with Higgs triplets and bitriplets}
\label{numass:triplet}
\qquad The relevant Yukawa couplings which leads to small non-zero neutrino
mass is given by
\begin{eqnarray}
{\cal L}^{II}_\nu &=& y_{ij} \ell_{iL} \Phi \ell_{jR}+ y^\prime_{ij} \ell_{iL}
\tilde{\Phi} \ell_{jR} +h.c.
\nonumber \\
&+& f^\prime_{ij}\ \left(\ell_{iR}^T \ C \ i \sigma_2 \Delta_R \ell_{jR}+
(R \leftrightarrow L)\right)+h.c.
\end{eqnarray}
The Majorana Yukawa couplings $f$ is same for both left and right handed neutrinos
because of left-right symmetry. After symmetry breaking, the effective mass matrix of
the neutrinos is
$$m_\nu=\frac{-f\,v^2\,v_R}{2\, m_\sigma\,s}-\frac{v^2}{v_R} y\, f^{-1}\,y^T=
m^{II}_\nu+m^{I}_\nu $$
Consider the values of y, f are of the order of unity, then the relative magnitude
of $m^{II}_\nu$ and $m^{I}_\nu$ depend on the parameters like $v_R$, $m_\sigma$, $s$. As discussed in section [\ref{sec2}], the type II term can become dominant (even if $v_R \sim 1$ TeV) if we take $m_{\sigma} \sim s \sim 10^8-10^{10}$ GeV.
\section{Results and Discussions}
\label{results}
\begin{itemize}
 \item Spontaneous breaking of Lorentz parity occurs via Higgs doublet in SUSYLR model with doublet Higgs only and via Higgs triplets/bitriplet in SUSYLR model with Higgs triplets and bitriplet. After taking into account of spontaneous D-parity breaking, the minimization of the scalar potential also allows the possibility of $M_R \sim \text{TeV},
v_L  \sim \text{eV}$  in LRSM with Higgs triplets and SUSYLR models with Higgs triplets and
Higgs bitriplet. It also allows $M_R \sim \text{TeV}, v_L/v_R \sim 10^{-6}$ in both Susy and
non-Susy LR models with Higgs doublets.
 \item In the SUSYLR model with Higgs doublets we can have a TeV scale $M_R$ as well as
$v_L/v_R \sim 10^{-6}$ by keeping the D-parity breaking scale very high  $\sim 10^{16}$ GeV.
The gauge couplings also unify for the same choice of scales although at the cost of adding extra particles which contribute to the beta functions at high energy. However if we add fermion triplets for seesaw, then unification is not possible with TeV scale $SU(2)_R$ breaking scale. Adding fermion singlet for seesaw purposes can evade this difficulty.
\item In SUSYLR model with Higgs triplet, the minimization conditions do not allow the
possibility of a TeV scale $M_R$ and eV scale $v_L$ simultaneously although gauge couplings
unify if we take $M_R$ as high as $10^{13}$ GeV. Thus we can not have TeV scale $M_R$,
type II seesaw dominance and gauge coupling unification simultaneously.
\item In SUSYLR model with Higgs triplets and bitriplet, we can have TeV scale $M_R$ and
eV scale $v_L$ only if we keep the D-parity breaking scale as low as $10^{10} $ GeV.
However such a choice of parity breaking scale spoils the gauge coupling unification.
The gauge couplings unify if we take $M_R = 10^{12}$ GeV and the D-parity breaking
scale as $10^{16}$ GeV with inclusion of two extra pairs of colored particles. Thus we can not have a TeV scale $M_R$ and unification simultaneously.
\end{itemize}

\section{Conclusion}
\label{conclusion}
In this work we have analyzed the different scenarios of spontaneous breaking of
D-Parity in both non-Susy and Susy version of left right symmetric models. We have
discussed the possibility of obtaining a TeV scale $M_R$, gauge coupling unification
and type II/type III seesaw dominance of neutrino mass within the
framework of different SUSYLR models. In all the  models where we explore the possibility of a TeV scale $M_R$, it is difficult to achieve unification with the minimal particle content. We have added some extra scalar particles as well as their superpartners with suitable transformation properties under the gauge group to achieve unification. We have shown that except for the SUSYLR model with Higgs doublets, we can not have a TeV scale $M_R$ and gauge coupling unification. In SUSYLR model with Higgs doublet, type III seesaw can dominate even if the D-parity breaking scale is as high as the GUT scale whereas in SUSYLR model with Higgs triplets and bitriplet, the D-parity breaking scale has to be kept as low as $10^{10}$ GeV for type II seesaw to dominate. However adding fermion triplets to give rise to seesaw spoils the unification with a TeV scale $M_R$ in the SUSYLR model with Higgs doublet. Adding fermion singlets instead of triplets do not give rise to this problem and can reproduce the necessary seesaw without affecting the RG evolution of the couplings.
\section{Acknowledgement}
Debasish Borah would like to acknowledge the hospitality of Physical Research Laboratory, Ahmedabad where this work was done.

\end{document}